# New Empirical Formula for (γ, n) reaction cross section near to GDR Peak for elements with Z ≥ 60

Rajnikant Makwana[1*], S. Mukherjee[1], Jian-Song Wang (王建松)[2], Zhi-Qiang Chen(陳志強)[2]

(1) Physics Department, Faulty of Science, The Maharaja Sayajirao University of Baroda, Vadodara-390002. India

(2) Key Laboratory of High Precision Nuclear Spectroscopy and Center for Nuclear Matter Science, Institute of Modern Physics, Chinese Academy of Science, Lanzhou 730000, China

(rajniipr@gmail.com)

**Abstract**

A new empirical formula has been developed that describes the (γ, n) nuclear reaction cross sections for the isotopes with Z ≥ 60. The results were supported by calculations using TALYS – 1.6, EMPIRE – 3.2.2 nuclear modular codes. The energy region for incident photon energy have been selected near to the giant dipole resonance (GDR) peak energy. The evaluated empirical data were compared with available data in the experimental data library - EXFOR. The data produced using TALYS – 1.6 and EMPIRE – 3.2.2 were in good agreement with experimental data. We have tested and presented the reproducibility of the present new empirical formula. We observed the reproducibility of the new empirical formula near to GDR peak energy is in good agreement with the experimental data and remarkable dependency on key nuclei properties: the neutron, proton and atomic number of the nuclei. The behavior of nucleus near GDR peak energy, and dependency of GDR peak on the isotopic nature were predicted. An effort has been made to explain the deformation of the GDR peak in (γ, n) nuclear reaction cross section for some isotopes, which could not reproduce with TALYS – 1.6 and EMPIRE – 3.2.2. The evaluated data have been presented for the isotopes $^{180}$W, $^{183}$W, $^{202}$Pb, $^{203}$Pb, $^{204}$Pb, $^{205}$Pb, $^{231}$Pa, $^{232}$U, $^{237}$U and $^{239}$Pu that have no previous measurements.

## 1. Introduction

Nuclear reactions are of prime importance in the application of nuclear reactor technology. Nuclear reactors (fusion-fission) require a complete dataset of neutron and photon induced reactions. Photonuclear reactions are becoming more important for the fusion reactors and accelerator driven sub-critical system (ADS), where high-energy photons will be generated and subsequently interact with the materials. The study of (γ, n) reactions are important for a variety of current and emerging fields, such as radiation shielding design, radiation transport, absorbed dose calculations for medical, physics, technology of fusion-fission reactors, nuclear transmutation and waste management applications [1,2]. In a fusion reactor, during the plasma shot, de-confined runaway electrons can interact with the first wall of the reactor that cause to produce high energy photons [3]. These high energy photons can open reaction channels like (γ, n), (γ, p), (γ, 2n), (γ, 3n), etc. The most prominent reaction is (γ, n), as it has a lower threshold than multi-neutron emission, whereas for charge particle emission Coulomb barrier needs to be considered. It is necessary to have exact information of the cross section for such nuclear reaction in order to perform accurate nuclear transport calculations. Tungsten (W) and beryllium (Be) are selected as first wall materials for the fusion reactor- International Thermonuclear Experimental Reactor (ITER) [4]. Among tungsten isotopes, only $^{182}$W (26.5 %), $^{184}$W (30.64%) and $^{186}$W (28.43%) have experimental cross section data for (γ, n) reaction. It is necessary to have the cross section of (γ, n) reaction for $^{180}$W (0.12%) and $^{183}$W (14.31%) along with all the remaining long-lived unstable isotopes, as they will interact with high-energy photons during the confined runaways and disruption phase [5]. Gamma induced nuclear reaction are also important for the nuclear transmutation (e.g. $^{234}$U(γ, n)$^{233}$U), which is useful for the nuclear safety and incineration. Importance of gamma incineration technique has been studied in the case of many isotopes for nuclear waste management [6-8].

In ADSs, the high energy proton beam will interact with high Z elements such as W, Pb-Bi, Th and U, which will produce neutrons through spallation reaction [9]. This spallation process will produce high energy photons, which will subsequently interact with the materials. It is necessary to have a complete nuclear dataset of photonuclear reaction for all isotopes of these elements. This can be done by experimental measurements. The experimental measurements of the nuclear reaction cross-section are one of the important method to complete the nuclear dataset. However, there are always limitations in the experimental measurements due to non-availability of all the energies of incident particles and preparation of target which may itself unstable. For complete nuclear data for several isotopes, nuclear modular codes such as TALYS – 1.6, EMPIRE – 3.2.2 are available. Using these codes one can predict the cross sections for different nuclear reaction channels. These codes basically use some nuclear models, and on the bases of the nuclear reaction theory, evaluation of the nuclear reaction data is done. The theory involved in photonuclear reaction cross section evaluation is discussed in the next section of the paper. Apart from this nuclear systematics and empirical formula provides alternative method for such isotopes, and it can efficiently predict the nuclear properties. Many authors have used this theoretical approach. There are several systematics and empirical studies already have been made for the photonuclear reactions [10]. These empirical formulae reduce experimental efforts, as they basically dependent on well-known nuclear properties. A new empirical formula has been developed and tested with nuclear modular codes and experimental data for $Z \geq 60$ in the present paper. With the help of the present empirical formula, one can predict the cross section datasets for those isotopes where there is a complete lack of the experimental data.

## 2. Theory of Photo Neutron Production

The interaction of high energy photons with target material can cause ejection of the nucleon/s, depending upon the energy of the incident photon. This reaction is considered as a photonuclear reaction. Photons should have sufficient energy above the binding energy of the nucleus for nucleon emission. As the nuclear binding energies are above 6 MeV for most of the isotopes, photons should have such threshold energy [11]. There are three basic mechanisms for the photonuclear reactions: (a) Giant dipole resonance (GDR), (b) Quasi-deuteron (QD) and (c) Intra-nuclear cascade [12]. A photon with energy below 30 MeV follows GDR mechanism. In this process, the photon energy is transferred to the nucleus by the oscillating electrical field of the photon, which induces oscillations among nucleons inside nucleus. The photo neutron production is more probable since proton ejection needs to overcome a large Coulomb barrier. For different isotopes at a particular energy, there is a peak of photo neutron production for (γ, n) reaction. This is called GDR peak energy. For isotopes above $Z = 60$, the peak energies are between 10-18 MeV. Above 30 MeV, the photo neutron production is mainly due to the QD effect [12]. In this mechanism, a photon interacts with the dipole moment of a pair of proton-neutron in place of nucleus as a whole [12]. Above 140 MeV, photo neutron production results from photo-pion production [12]. Further the study of thermal fluctuation on GDR parameters are also of interest and studies are ongoing [13-16].

## 3. Present Empirical Formula and Theoretical Calculations

According to semi classical theory of the interaction of photons with nuclei, the shape of the fundamental resonance of the photo absorption cross section follows a Lorentz curve [12, 17].

$$\sigma(E) = \frac{\sigma_i}{1 + [\frac{(E_\gamma^2 - E_m^2)^2}{E_\gamma^2 \gamma^2}]} \quad (1)$$

Where, $\sigma_i$, $E_\gamma$ and $\gamma$ are the Lorentz parameters: peak cross section, resonance energy and full width at half maximum respectively [19].

In a more general way, by using nuclear modular codes, such as TALYS – 1.6 and EMPIRE – 3.2.2, the photo absorption cross section is calculated as the sum of two components [20],

$$\sigma_{abs}(E_\gamma) = \sigma_{GDR}(E_\gamma) + \sigma_{QD}(E_\gamma) \quad (2)$$

The component $\sigma_{GDR}(E_\gamma)$ represents GDR and is given by Lorentzian shape, which describes the giant dipole resonance. It is given from the eq. (1) by following expression,

$$\sigma(E) = \sum_i \frac{\sigma_i \cdot (E_\gamma \cdot \Gamma_i)^2}{(E_\gamma^2 - E_i^2)^2 + (E_\gamma \cdot \Gamma_i)^2} \quad (3)$$

Where $\sigma_i$, $E_i$ and $\Gamma_i$ are: peak cross section, resonance energy and full width at half maximum respectively. The summation is limited to i=1 for spherical nuclei, while for deformed nuclei the resonance is split and one uses i=1, 2. The component $\sigma_{QD}(E_\gamma)$, is given by Levinger type theory given by Chadwick et al. [20 - 22]. It is basically from the quasi deuteron model. In the energy range photonuclear threshold to 30 MeV, the GDR mechanism is dominant, 30 – 140 MeV QD mechanism is dominant. Above 140 MeV the threshold energy for pion production is achieved [21].

Above theory has been used in TALYS – 1.6 and EMPIRE – 3.2.2 nuclear modular codes [23, 24]. Further details of these codes are given in references 18 – 19. Using these codes, (γ, n) nuclear reaction cross section for different isotopes ($Z \geq 60$) were calculated and presented in present work. Till now the photonuclear reaction cross sections are evaluated using the Lorentz parameters. These parameters for several isotopes are calculated by fitting the experimental data or by systematics [25].

### 3.1. Fundamental Term

In the present paper, in contrast to the Lorentzian parameters, the basic properties of nuclei A, N and Z are used to estimate photonuclear cross section. Levovskii had given empirical formulas for (n, p) and (n, 2n) reaction cross section at 14.0 MeV [26],

$$\sigma(n, p) \propto \sigma_p \cdot e^{-\frac{33 \cdot (N-Z)}{A}} \tag{4}$$

$$\sigma(n, 2n) \propto \sigma_\alpha \cdot e^{-\frac{33 \cdot (N-Z)}{A}} \tag{5}$$

Where, $\sigma_p = \pi r_0^2 (A^{1/3} + 1)^2$ and $\sigma_\alpha = 0.4 \cdot \pi r_0^2 (A^{1/3} + 1)^2$

$r_0 = 1.2 \times 10^{-13}$ cm

These empirical formulae are based on A, N and Z of a nucleus, and at an energy 14.0 MeV. Similarly, it is possible to derive an empirical formula for photo induced (γ, n) nuclear reaction, which may be applied near to GDR peak energy. For the (γ, n) reaction, the formula is modified in the following way,

$$\sigma(\gamma, n) \propto \sigma_m \cdot e^{-\frac{33.5 \cdot (N-Z)}{A}} \tag{6}$$

$$\sigma_m = \pi r_0^2 \cdot (A^{2/3} + 1)^2 \cdot (N - Z) \cdot A^{-\frac{4}{3}} \tag{7}$$

Where $r_0$ = average nuclear radius

### 3.2. Isotopic Dependent Resonance Term

The term (N – Z)/A is the asymmetry parameter which considers the deformation of a nucleus. In this expression, there is no term containing energy dependency. Hence, an energy dependent term must be added, and the modified formula is as given below,

$$\sigma(\gamma, n) \propto \sigma_m \cdot e^{-\frac{33.5 \cdot (N-Z)}{A}} \cdot e^{\left(-\left(\frac{E_i - S_j \cdot R_p}{2}\right)^2\right)} \quad (8)$$

Where $E_i$ is the incident photon energy, $R_p$ = Resonance parameter.

The parameter $S_j$ is given by,

$$S_j = \frac{A^2}{2(N-Z)^2} \quad (9)$$

The parameter $R_p$ is estimated for an isotope, by fitting the (γ, n) nuclear reaction cross section using the above formula for different isotopes of the same element. We observed that this parameter $R_p$ is following a linear relationship against the atomic mass of the different isotopes of the same element, which can be written in the form of the following equation,

$$R_p = m \cdot A + C \quad (10)$$

Where A is the atomic mass of the isotope, m and C are slope and intercepts respectively, for different elements are discussed in subsection 3.1. The detail of this parameter ($R_p$) is given in the subsection 3.1.

This term $e^{-[\{\frac{(E_i - S_j \cdot R_p)}{2}\}^2]}$ depends on the energy of the incident photon and the isotopic nature of the target nucleus. When a photon incident on the nucleus, the response of the nucleus depends on the photon energy. Until the incident photon do not have threshold energy of the photo fission, the photon cannot eject a nucleon from the nucleus. If energy of the photon is above the threshold energy of the (γ, n) reaction, the probability of the reaction, cross section increases until the resonance peak energy. After this energy, cross section decreases again. This is incorporated using this exponential term. The subtraction of $S_j \cdot R_p$ from the incident photon energy shows the isotopic dependence of the resonance peak energy of the reaction. As the isotopic number increases it is observed in the experimental data that the GDR peak is shifting towards lower energy side. This back shift effect can be calculated with the exponential term considered here. The value of $S_j \cdot R_p$ increases with addition of neutron to the isotope nucleus. This means that when photon incidents on the target isotope, it interacts with the last shell neutron in the nucleus. The binding energy of the last added neutron will be least. Hence photon may require smaller energy for occurring the resonance as the isotope number increases.

This can be observed from the following Fig-1-2, showing isotopic effect for the resonance peak energy back shifting in Nd and Pt isotopes from the above exponential term.

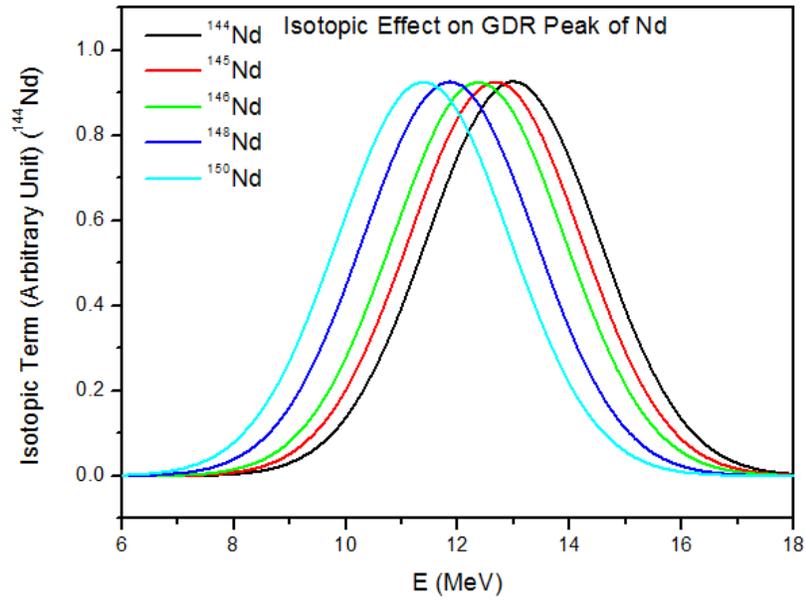

**Figure 1:** Back shift of Resonance Peak Energy in Nd isotopes which is result from the term $e^{\left(-\left(\frac{Ei - S_j \cdot R_p}{2}\right)^2\right)}$

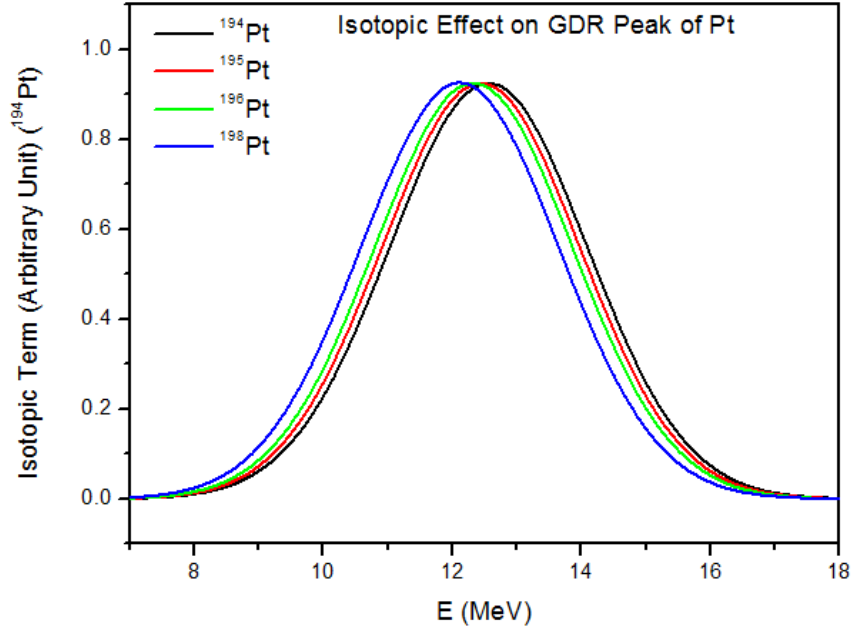

**Figure 2:** Backshift of Resonance Peak Energy in Pt isotopes which is result from the term $e^{\left(-(\frac{E_i - S_j \cdot R_p}{2})^2\right)}$

### 3.3 Energy Dependency Term

It was found that another energy related term is required to make the formula more efficient to predict the cross section. If the photon energy increases, then the photon can transfer more energy to the nucleus. In the mechanism of GDR, the oscillating electrical field transfers its energy to nucleus by inducing an oscillation in nucleus, which leads to relative displacement if tightly bound neutrons and protons inside the nucleus. [12]

When the energy of the photon is low (near to threshold), the oscillating electric field of the photon interacts with the collective nucleus field produced by the sum effect of nucleons. But as the energy of the photon increases, the oscillating electrical field interacts with a pair of neutron and proton rather than the nucleus. This is followed by the term $e^{\sqrt{1+E^{\frac{2}{3}}}}$. where E = energy of the incident photon. This term shows that photon can have more energy to transfer the nucleon as the incident photon energy increases. It indicates that as the energy of the photon increases, it can have less interaction time with nucleons, and hence the pre-equilibrium or direct reaction mechanism can be followed for the emission of the neutron.

Hence, by the addition of an energy dependent term, the modified formula is,

$$\sigma(\gamma, n) \propto \sigma_m \cdot e^{-\frac{33.5 \cdot (N-Z)}{A}} \cdot e^{\left(-\left(\frac{(E_i - S_j \cdot R_p)}{2}\right)^2\right)} \cdot e^{\sqrt{1+E^{\frac{2}{3}}}} \quad (11)$$

An additional factor $S_f$ which is an isospin dependent factor, has been introduced to complete the formula. This factor was plotted for different isotopes of a same element, and fitted. We observed this factor follows some exponential relation, which is described in subsection 3.2. This empirical formula gives the cross section in order of mili-barn.

The final modified formula is now,

$$\sigma(\gamma, n) = \sigma_m \cdot e^{-\frac{33.5 \cdot (N-Z)}{A}} \cdot e^{\left(-\left(\frac{(E_i - S_j \cdot R_p)}{2}\right)^2\right)} \cdot e^{\sqrt{1+E^{\frac{2}{3}}}} \cdot S_f \quad (12)$$

### 3.3.1     $R_p$ Parameter

This parameter shows the dependency of the photo absorption cross section with respect to the atomic number (Z). In the empirical formula the term $e^{-[\{\frac{(E_i - S_j \cdot R_p)}{2}\}^2]}$, the subtraction factor contains $S_j$ and $R_p$. The $R_p$ is responsible for the change in the cross section due to atomic number where the multiplication of $S_j$ and $R_p$ is responsible for the isotopic back shift effect, which is shown in fig.-1- 2. The parameter $R_p$ for different isotopes can be calculated using a linear relation viz eq. (10) with the atomic mass number of isotopes for an element. Therefore, the plot of $R_p$ vs A for different elements should parallel lines with different intercepts on $R_p$ axis as shown in Fig.3. It is a property of parallel lines that they have same slope but different intercepts. Hence, the mean slope of different element has been taken as the standard slope for all elements ($Z \geq 60$). This value of the slope (m) mentioned in eq. (10) is ~ 0.03164 ± 0.00409. The intercept C for different elements are plotted against the atomic number of an element, and fitted with mathematical software MATLAB, using 3$^{rd}$ degree polynomial as shown in Fig.4. The intercept C for different element can be determined from the following equation.

Linear model Polynomial 3$^{rd}$ degree:

$$C(Z) = p_1 \cdot Z^3 + p_2 \cdot Z^2 + p_3 \cdot Z + p_4 \quad (13)$$

$p_1 = -4.155 \times 10^{-5}$, $p_2 = 0.008971$, $p_3 = -0.7156$, $p_4 = 15.78$, Z = Atomic number

(SSE: 0.00147; R-square: 0.9998: Adjusted R-square: 0.9996; RMSE: 0.01917)

Hence, the intercept for any element can be evaluated using the above eq. (13). Using this intercept and the slope 0.03164 ± 0.00409 one can calculate the parameter $R_p$ from eq. (10). The model values of the parameter $R_p$ for different elements are compared with the previous manually selected values are compared in Fig.3.

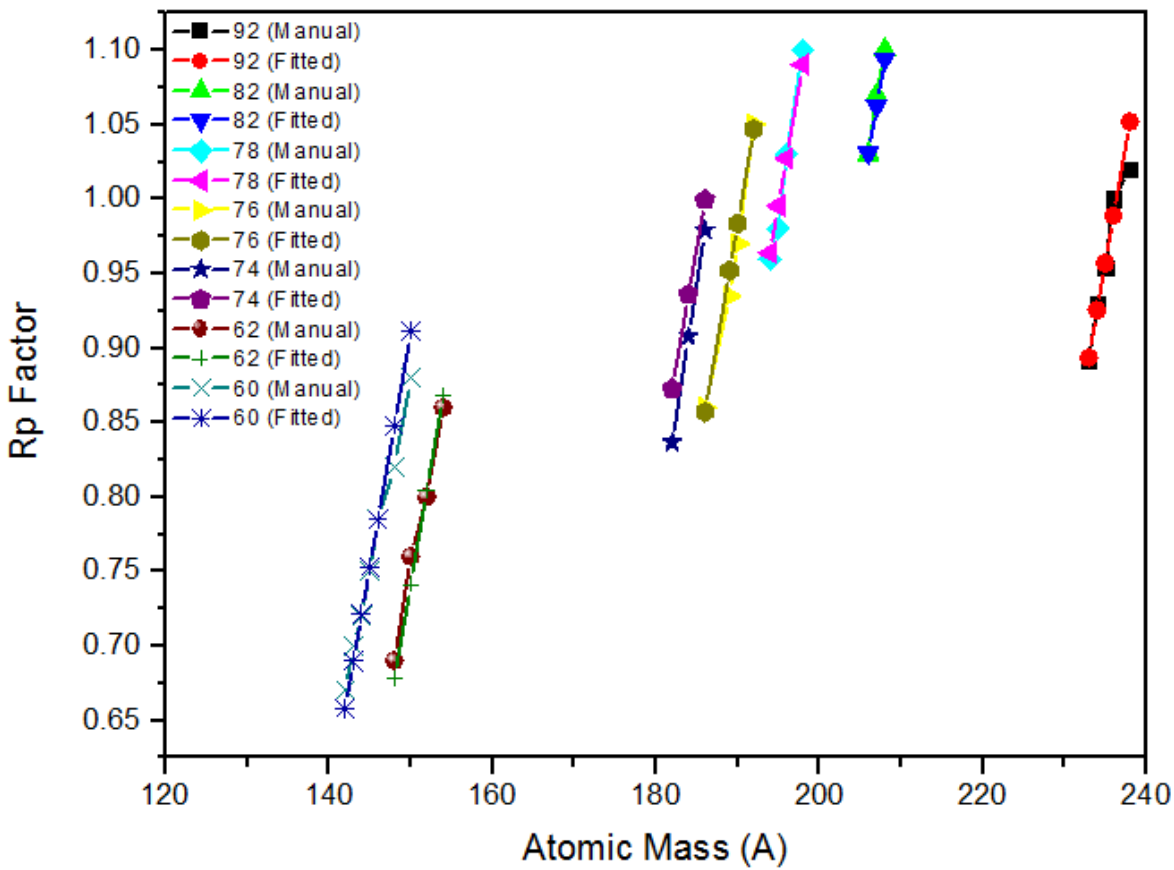

**Figure 3:** $R_p$ parameter fitting for different elements with eq. (10)

### 3.3.2 $S_f$ Parameter

This parameter includes the isospin effect. This effect has been discussed by J. S. Wang et al. [27]. In order to include this effect in the empirical formula, an additional factor called $S_f$ has been added. This factor was initially manually added and then, in order to generalize, it is fitted with different combinations of N, Z and A. It was found that it follows a complex exponential relation with $\exp((N-Z)/N)$ of an isotope. This parameter $S_f$ is also considered as a result of the asymmetry of the nucleus. As there is a difference in neutron and proton number, the fraction (N – Z)/N is the available neutron fraction for a photon to eject. As this fraction value increases, the value of $S_f$ increases respectively, which directly shows increment in the photo absorption cross section of that isotope.

This isotopic factor $S_f$ for different isotopes is plotted with respect to $e^{\frac{(N-Z)}{N}}$ and fitted with MATLAB software as shown in Fig. 4. The generalized expression to determine $S_f$ parameter for an isotope is as given below.

$$S_f = ae^{bx} + ce^{dx} \qquad (14)$$

Where, x = (N – Z)/N, a = 1.21 × 10$^{-22}$, b = 34.21, c = 7.71 × 10$^{-11}$, d = 14.52

( SSE: 0.006977; R-square: 0.9781; Adjusted R-square: 0.9759; RMSE: 0.01551)

If we observe fig.-5 carefully, some points when $e^{\frac{(N-Z)}{N}}$ is near to 1.40 to 1.42, they have almost same $S_f$ factor values. These $S_f$ factor values are for Z = 82 and N = 124, 125, 126, which are either a magic number or near to the magic number. $S_f$ is purely depending on (N – Z)/N, which is a shell depended term. The anomalous behavior of same $S_f$ factor values for these isotopes is because of the magic shell effect.

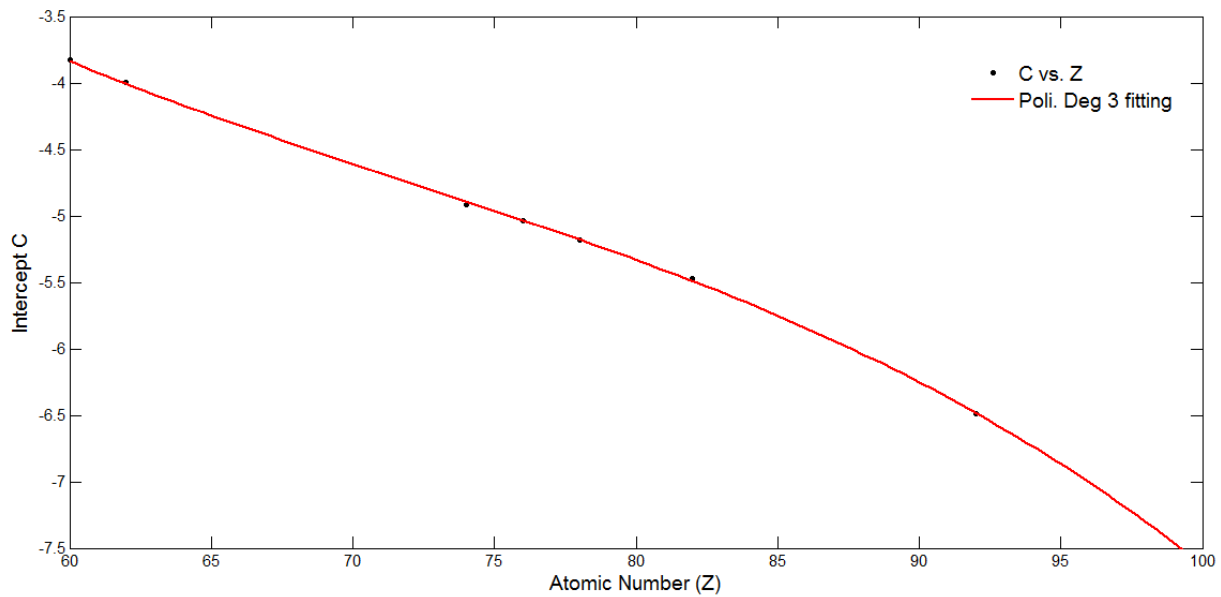

**Figure 4:** Intercept C for eq. (10) for different elements fitted with eq. (11)

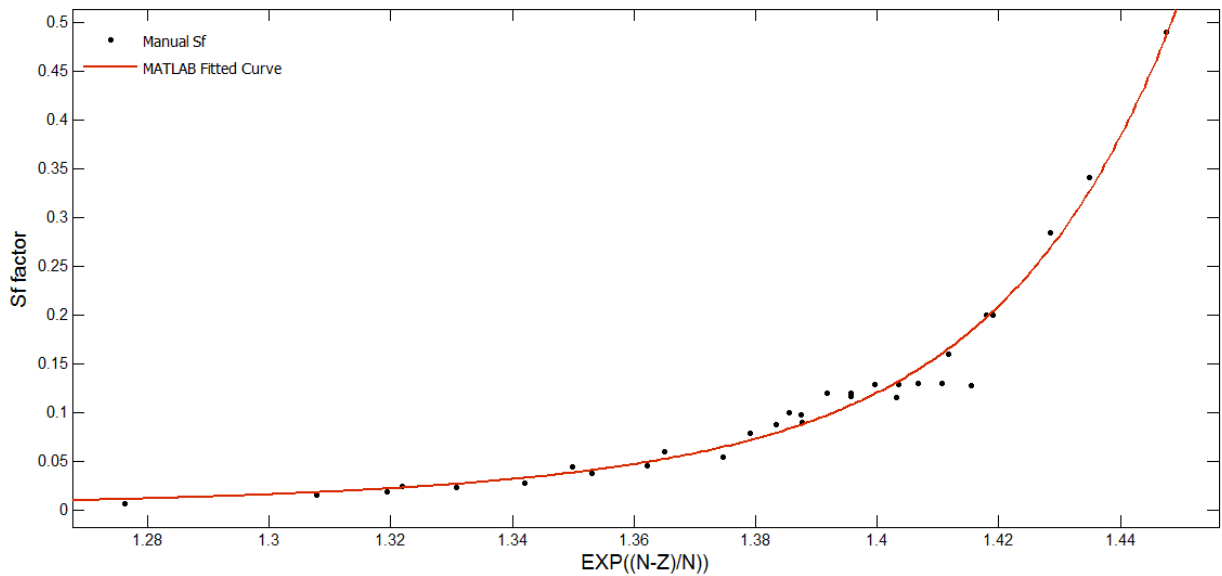

**Figure 5:** $S_f$ parameter for different (N-Z)/N fitted with eq. (14)

## 4  Results and Discussion

The (γ, n) reaction cross sections are calculated using TALYS – 1.6, EMPIRE – 3.2.2 and newly developed empirical formula for the isotopes with Z ≥ 60 and presented in Figs.6 – 10. The cross sections are calculated for the energy range in which GDR peak is observed. The theoretically calculated cross sections are compared with the experimentally available data in EXFOR data library [28]. The data calculated using modular codes and empirical formula are in agreement with the experimental data as shown in the Figs.6 – 10. However, the cross section values and the nuclear behavior near to the GDR peak predicted by the empirical formula are more appropriate. This empirical formula is good for those isotopes having a single GDR peak. In most of the cases studied here, that have a single GDR peak, the empirical formula is producing good agreement near to the GDR peak energy compared to the Lorentz Curve fitting based model.

In the case of the isotopes with Z from 63 to 75, it is found that according to collective model these isotopes have large nuclear quadrupole moment. The quadrupole exists because of the asymmetry of the nucleus. The nuclei are found in the middle of 1d, 2s shells in the range of 145 < A <185. The energy difference between the ground state and the first excited state is in order of 100s of keV. In the deformed nucleus the incident photon can interact either with the ground state or with the excited state nucleon. And hence it can produce resonance at two different nearby energies. Which is observed in the above isotopes. For such case Lorentz cur3333ve based model viz. TALYS – 1.6 and EMPIRE – 3.2.2 is reliably working for these isotopes shown in Fig.11. But for some cases this TALYS – 1.6 and EMPIRE – 3.2.2 model is also not working well e.g. Figs. 11(e – f). To apply the empirical formula for such isotopes it is assumed that, there may be two peaks due to unresolved resonances occurring nearby energies of ground and excited nuclei, which are due to quadrupole moment. This suggests parameters $R_p$ and $S_f$ can have two different values for these isotopes. It indicates that the energy dependence cross section curve is made of two curves with two different $R_p$ ($R_{p1}$ and $R_{p2}$) and $S_f$ values ($S_{f1}$ and $S_{f2}$) of parameters $R_p$ and $S_f$ respectively. These values can be estimated by multiplying following factors to the $R_p$ and $S_f$ values calculated from section 3.1 and 3.2.

$$R_{p1} = 0.95 \times R_p \tag{15}$$

$$R_{p2} = 1.20 \times R_{p1} \tag{16}$$

$$S_{f1} = 1.39 \times S_f \tag{17}$$

$$S_{f2} = 0.28 \times S_{f1} \tag{18}$$

The two curves are intersecting at a deep point, where both curves should have the same value of cross section. This intersection point energy can be calculated by comparing the right side of the eq. (12) for above values.

$$\sigma_m \cdot e^{-\frac{33.5 \cdot (N-Z)}{A}} \cdot e^{\left(-\left(\frac{(E_i - S_j \cdot R_{p1})}{2}\right)^2\right)} \cdot e^{\sqrt{1+E^{\frac{2}{3}}}} \cdot S_{f1} = \sigma_m \cdot e^{-\frac{33.5 \cdot (N-Z)}{A}} \cdot e^{\left(-\left(\frac{(E_i - S_j \cdot R_{p2})}{2}\right)^2\right)} \cdot e^{\sqrt{1+E^{\frac{2}{3}}}} \cdot S_{f2} \tag{19}$$

Solving this eq. we get,

$$E_{deep} = \frac{1}{2} S_j \cdot (R_{p1} + R_{p2}) + \frac{2 \ln\left(\frac{S_{f2}}{S_{f1}}\right)}{S_j (R_{p1} - R_{p2})} \tag{20}$$

This energy $E_{deep}$ is near to the threshold energy of the (γ, 2n) reaction. With this consideration the results are plotted in fig.11(a – f).

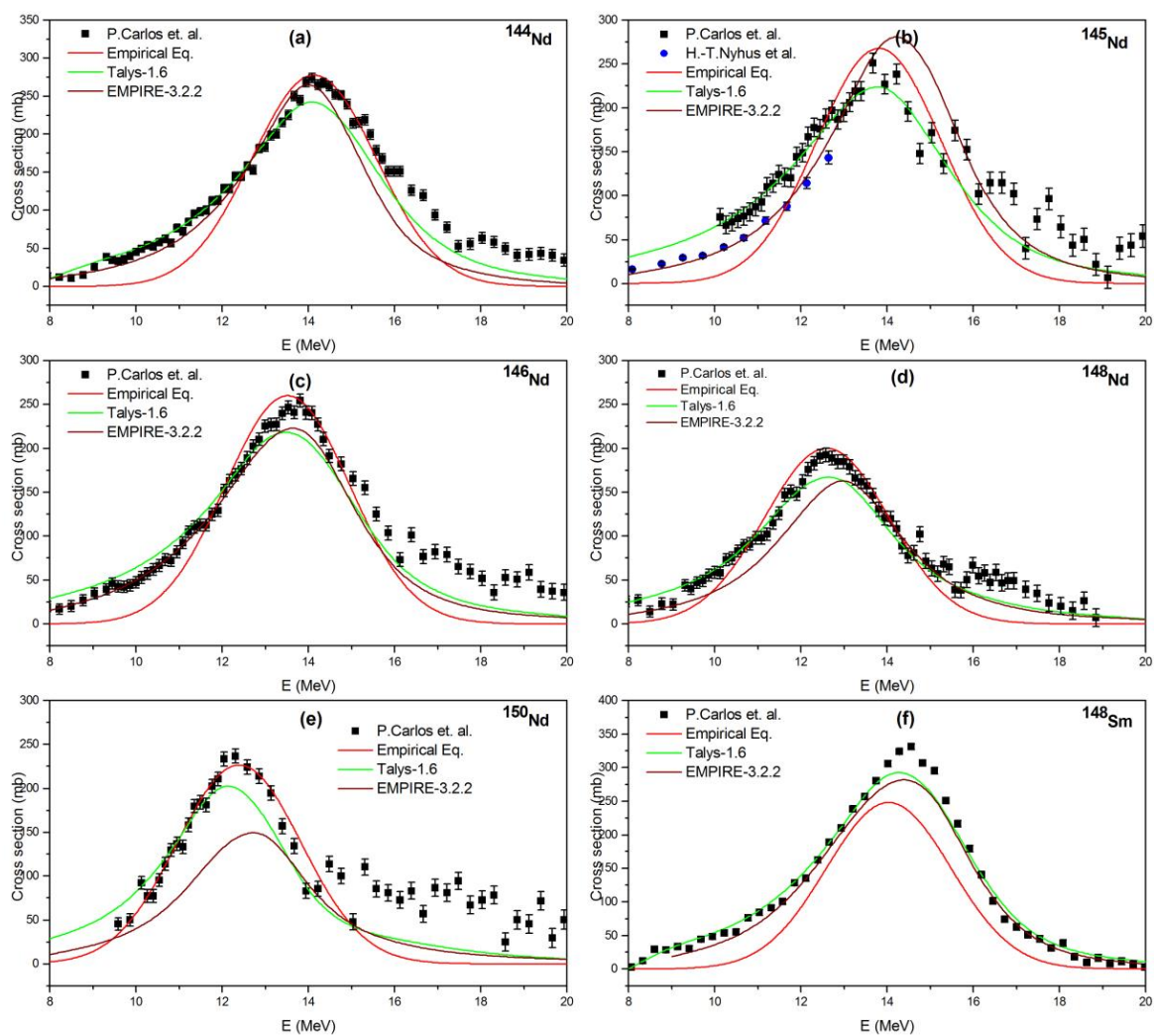

**Figure 6** Comparison of Evaluated data using TALYS-1.6, EMPIRE-3.2.2, and Empirical Formula with Experimental data from EXFOR comparison for $^{144}$Nd, $^{145}$Nd, $^{146}$Nd, $^{148}$Nd, $^{150}$Nd, $^{148}$Sm

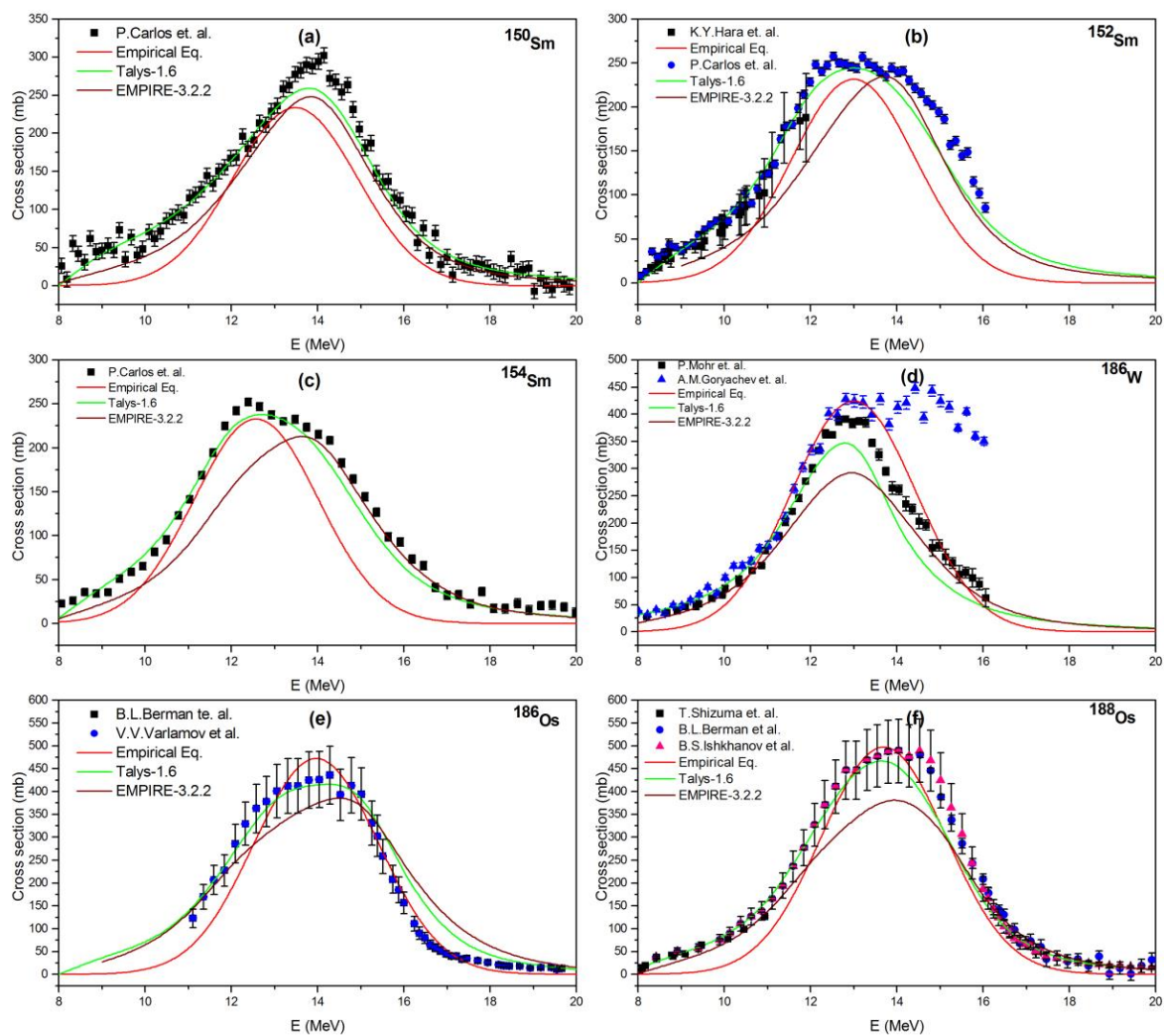

**Figure 7** Comparison of Evaluated data using TALYS-1.6, EMPIRE-3.2.2, and Empirical Formula with Experimental data from EXFOR comparison for $^{150}$Sm, $^{152}$Sm, $^{154}$Sm, $^{186}$W, $^{186}$Os, $^{188}$Os

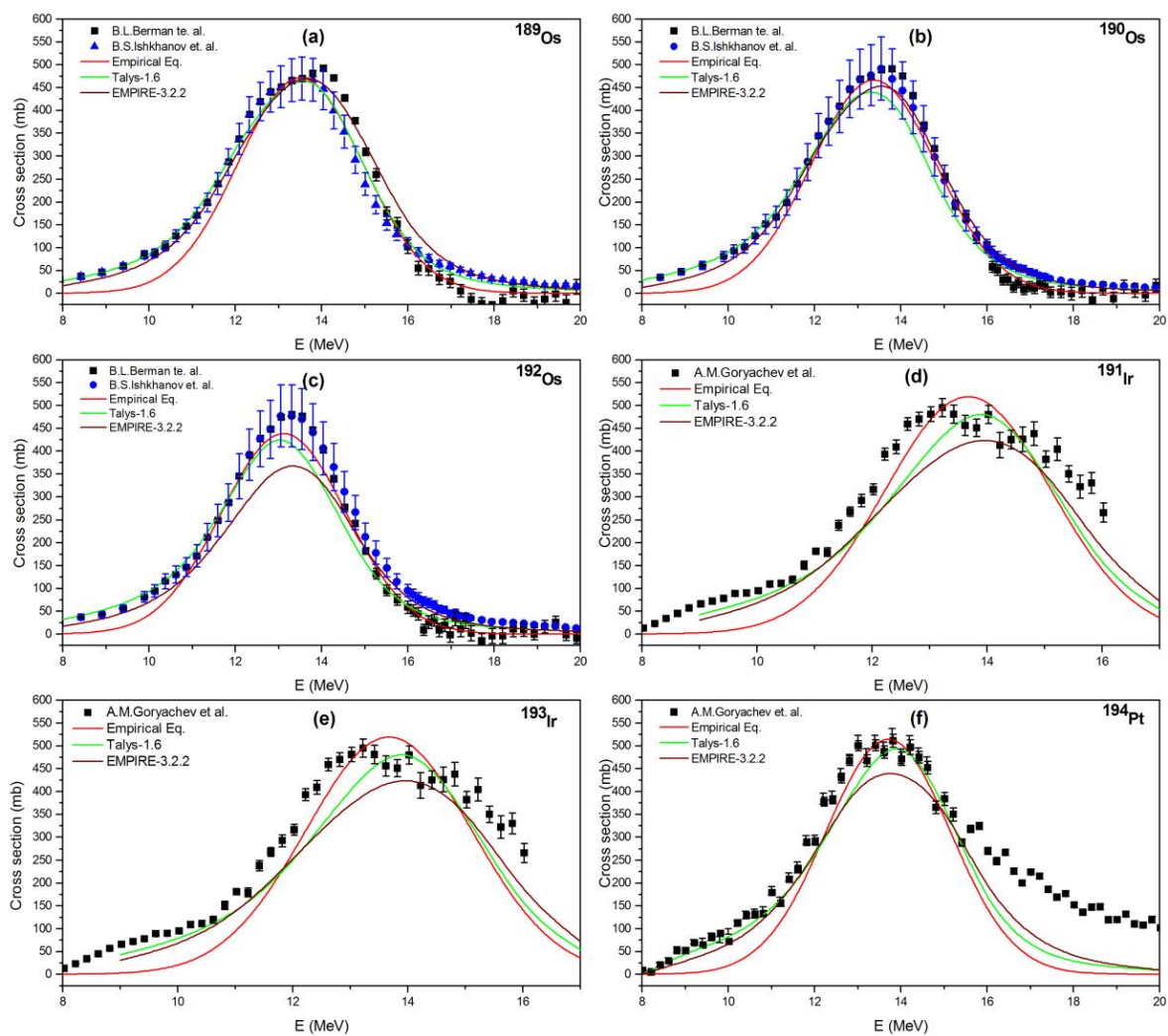

**Figure 8** Comparison of Evaluated data using TALYS-1.6, EMPIRE-3.2.2, and Empirical Formula with Experimental data from EXFOR comparison for $^{189}$Os, $^{190}$Os, $^{192}$Os, $^{191}$Ir, $^{193}$Ir, $^{194}$Pt

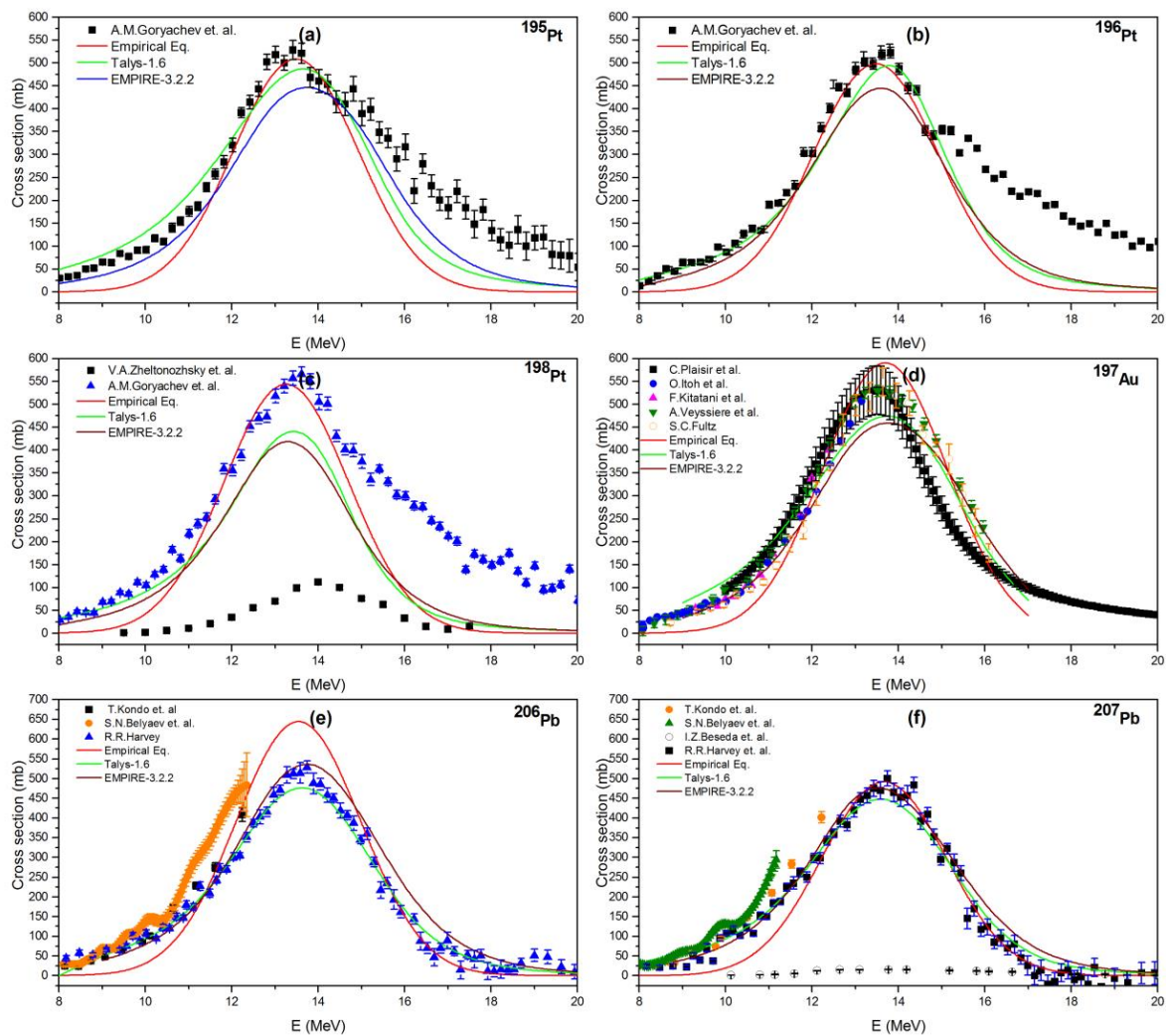

**Figure 9** Comparison of Evaluated data using TALYS-1.6, EMPIRE-3.2.2, and Empirical Formula with Experimental data from EXFOR comparison for $^{195}$Pt, $^{196}$Pt, $^{198}$Pt, $^{197}$Au, $^{206}$Pb, $^{207}$Pb

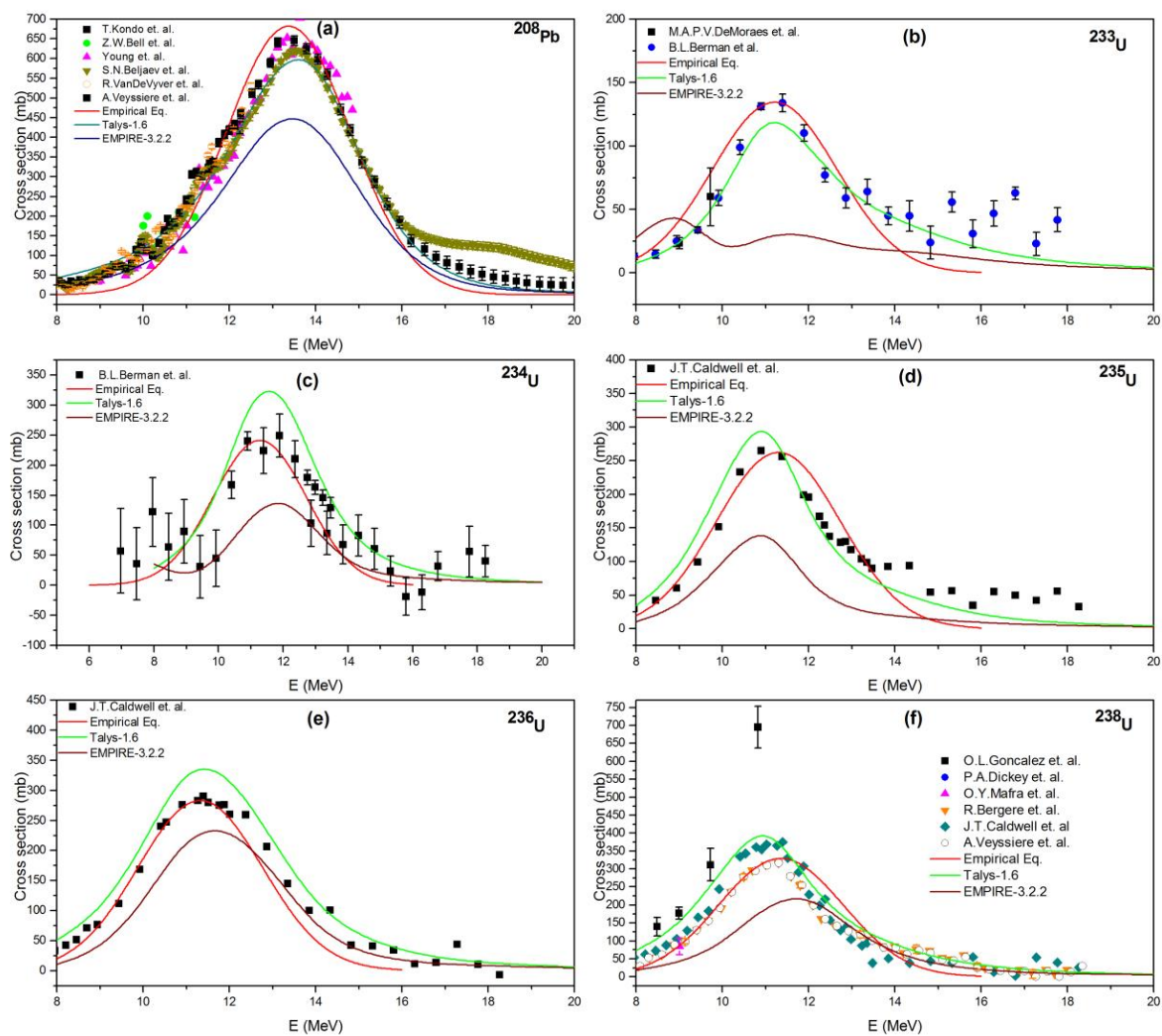

**Figure 10** Comparison of Evaluated data using TALYS-1.6, EMPIRE-3.2.2, and Empirical Formula with Experimental data from EXFOR comparison for $^{208}$Pb, $^{233}$U, $^{234}$U, $^{235}$U, $^{236}$U, $^{238}$U

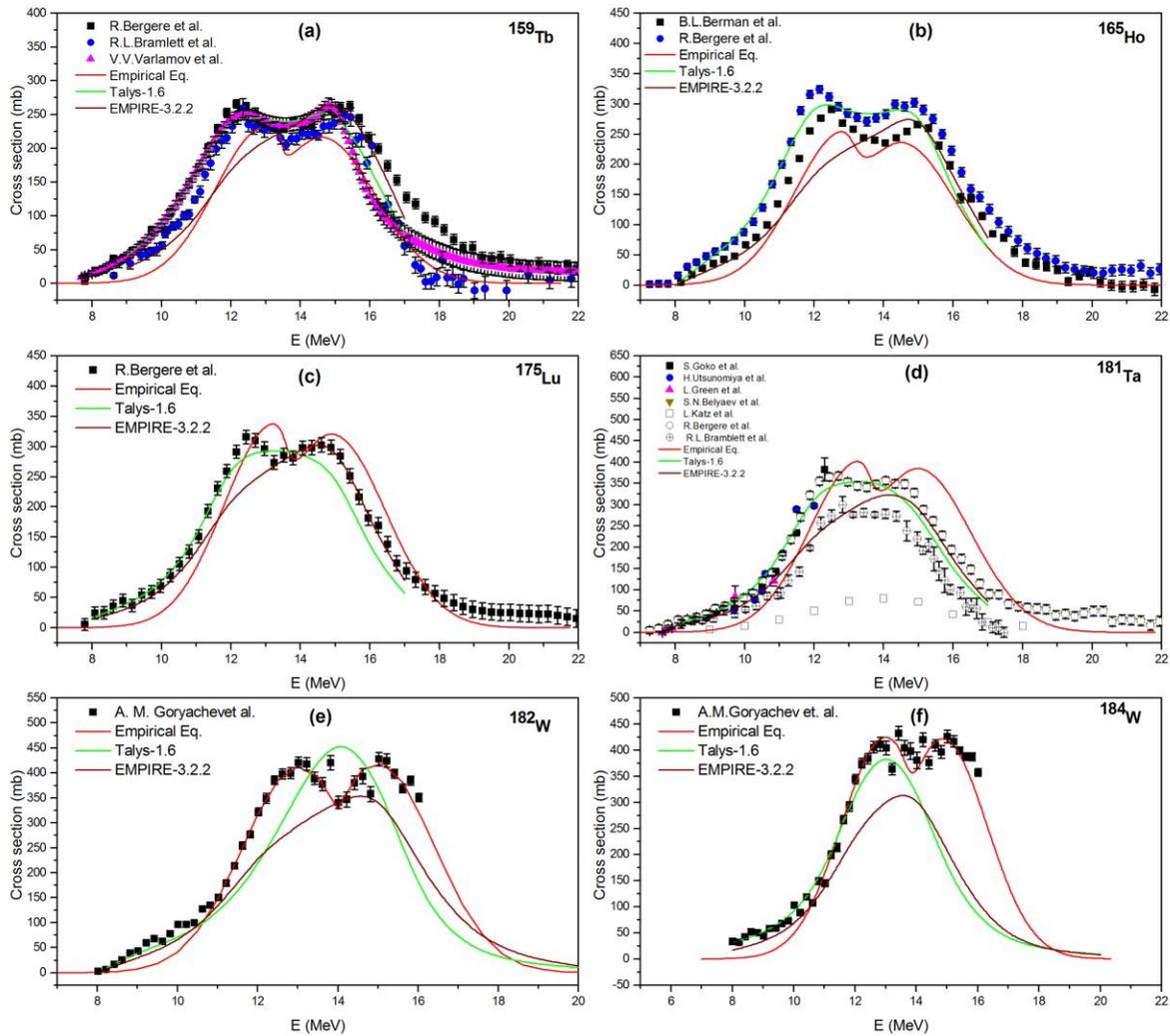

**Figure 11** Effect of a deformed nuclei in (γ, n) nuclear reaction, data comparisons for TALYS – 1.6, EMPIRE – 3.2.2 and Present Empirical formula

## 5 Applications

The (γ, n) cross section for several isotopes of W, Pb, Pa, U and Pu, which have no available experimental data, are calculated and presented using TALYS – 1.6, EMPIRE – 3.2.2 and present Empirical formula. Further the predicted data of the isotopes were compared with different standard evaluated data libraries, wherever available.

Tungsten is a prime candidate for plasma facing component in a fusion reactor. It is selected for the diverter material in ITER fusion reactor [4]. Tungsten isotopes $^{182}$W, $^{184}$W and $^{186}$W have available experimental data for the (γ, n) reaction cross section [28]. The (γ, n) cross section for remaining isotopes $^{180}$W(0.12%) and $^{183}$W(14.31%) are calculated and compared with the evaluated data available in ENDF/B-VII.1. No other standard data library is having photonuclear data for these tungsten isotopes [19]. There is an agreement between present evaluated data and ENDF/B-VII.1 as can be seen in Fig. 12 (a – b) data. Lead is a prime element of the Pb-Li blanket module of the fusion reactor as well as, it is also a candidate of the ADS target material [29]. Lead isotopes $^{206}$Pb, $^{207}$Pb and $^{208}$Pb have available experimental data. The (γ, n) cross section for remaining isotopes of lead $^{202}$Pb ($5.25 \times 10^4$ y, [30]), $^{203}$Pb (51.92 h, [30]), $^{204}$Pb ($1.4 \times 10^{17}$ y, [30]) and $^{205}$Pb ($1.73 \times 10^7$ y, [30]) are calculated and presented. These isotopes of lead have large half-life and they are facing high energy photons during the runaway electron generation and the disruption phase in plasma [5]. There are some isotopes of Pa and U: $^{231}$Pa($3.27 \times 10^4$ y, [30]), $^{232}$U(68.9 y, [30]) and $^{237}$U(6.75 d, [30]) having no evaluated cross section data available in different standard data libraries, such as ENDF/B-VII.1, JENDL-4.0, JEFF-3.1, ROSFOND and CENDL-3.1 [31,32] are also calculated and presented here. The evaluated data for $^{239}$Pu($2.41 \times 10^4$ y, [30]) and available data in ENDFB/VII.1 are presented in Fig. 13 (d). Though in present context, cross sections are evaluated for limited isotopes, it can be applied to calculate (γ, n) reaction cross section for actinides using the nuclear modular codes and present empirical formula. Further the TALYS – 1.6 and EMPIRE – 3.2.2 codes can be used to calculate the (γ, n) reaction cross section for the isotopes, which have available GDR parameters, whereas the present empirical formula can be used to calculate cross section for any isotope with $Z \geq 60$.

Another important application is, by using the nuclear modular codes and present formula, it is possible to calculate the incident gamma energy for which the cross section will have maximum value, i.e. the GDR peak energy. It can be used to calculate the incident charge particle (e.g. electron) beam energy for the bremsstrahlung production, which is required to design a photo neutron source. There are some theoretical transport codes available to transport the electrons and photons such as MCNP [12, 33-34], FLUKA [35, 36], GEANT [37] etc. With these codes, one can estimate the bremsstrahlung spectra from the electron beam.

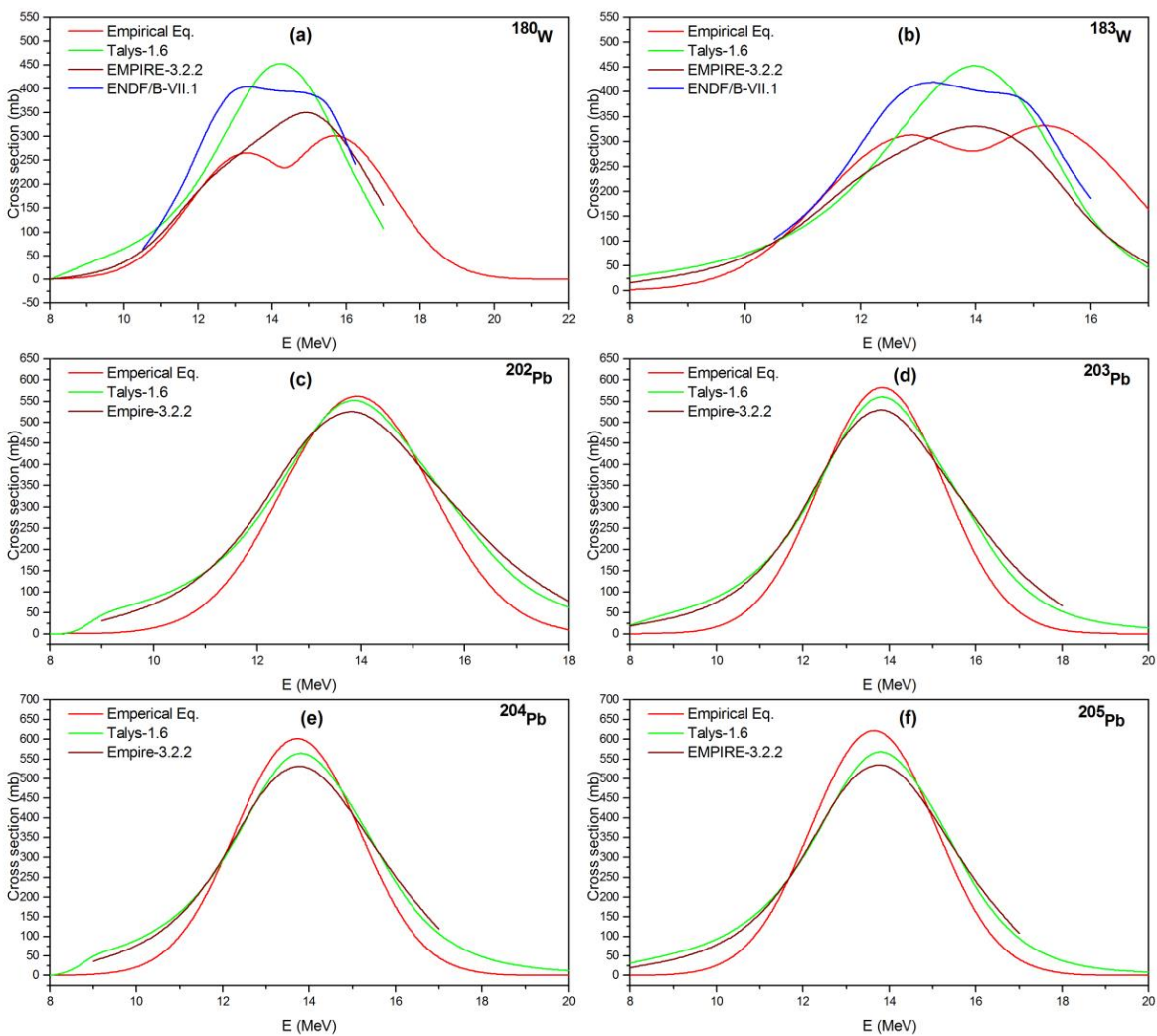

**Figure 12** Comparison of Evaluated data for $^{180}$W, $^{183}$W, $^{202}$Pb, $^{203}$Pb, $^{204}$Pb, $^{205}$Pb using TALYS -1.6, EMPIRE-3.2.2, and Empirical Formula

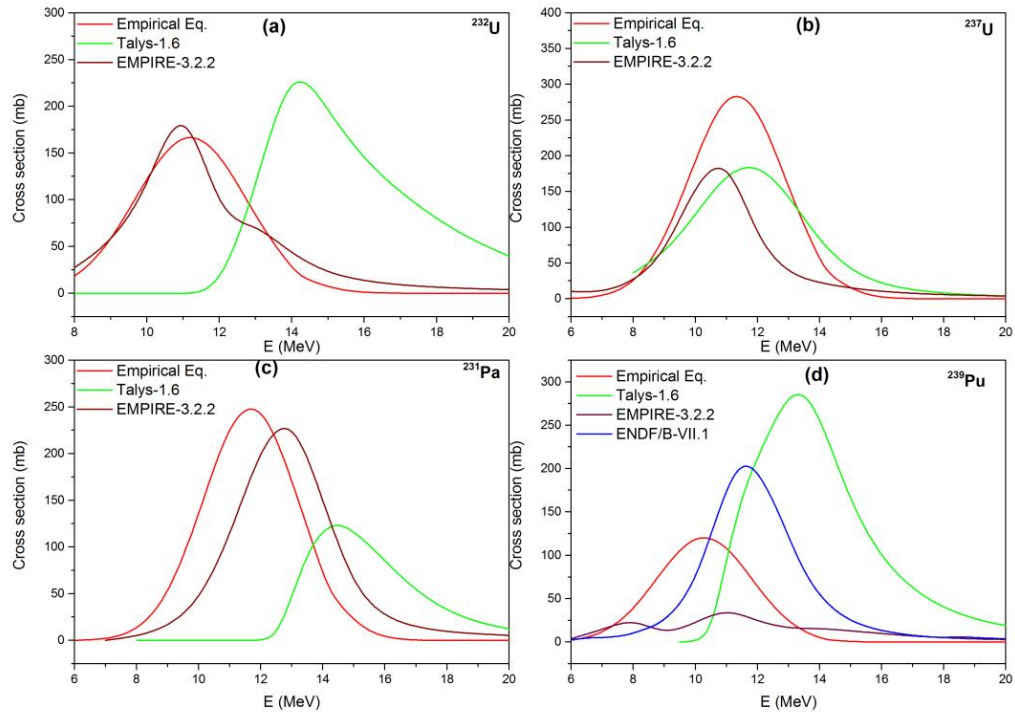

**Figure 13** Comparison of Evaluated data for $^{231}$Pa, $^{232}$U, $^{237}$U, $^{239}$Pu using TALYS -1.6, EMPIRE-3.2.2, and Empirical Formula

## 6   Conclusion

In the present work, a new empirical formula has been developed to investigate the (γ, n) reaction cross section for different isotopes with Z ≥ 60 in the GDR energy region. The results for the (γ, n) reaction cross section obtained by using the above empirical formula have been reproduced by using nuclear modular codes: TALYS – 1.6 and EMPIRE – 3.2.2. It was shown that TALYS – 1.6, EMPIRE – 3.2.2 and empirical formula are in agreement with the experimental data. Further a conclusion may be drawn that there may be no deformation in the GDR peak of a pure (γ, n) reaction cross section for spherical nucleus. As a result of the quadrupole, which is due to asymmetric shape of nucleus, the present deformation has been observed.

In addition to this, the evaluated data for $^{180}$W, $^{183}$W, $^{202}$Pb, $^{203}$Pb, $^{204}$Pb, $^{205}$Pb, $^{231}$Pa, $^{232}$U, $^{237}$U and $^{239}$Pu were presented using TALYS – 1.6, EMPIRE – 3.2.2 and Empirical formula. Among these only $^{180}$W, $^{183}$W and $^{239}$Pu have evaluated data in ENDF/B-VII.1 [29], which are compared with the present evaluated data. For $^{180}$W and $^{183}$W, present evaluated data are in good agreement, but in the case of $^{239}$Pu, it is in contrast. It is necessary to do experiments in the GDR energy range to validate the present evaluated data for $^{239}$Pu. Further, though here only limited isotopes are used for the (γ, n) reaction cross section evaluation, the trend used in the paper may be useful for other isotopes provided Z ≥ 60 for Empirical formula.

# 7    Acknowledgement

Intensive discussions with M. Herman (NNDC, Brookhaven National Laboratory, USA), R. Kapote (NAPC-Nucear Data Section, IAEA, Austria), P. K. Mehta (The M. S. University of Baroda, Vadodara), P. Mishra (The M. S. University of Baroda, Vadodara), N. Agrawal (The M. S. University of Baroda, Vadodara) are graceful to acknowledge.

# 8    References:

[1] http://www-pub.iaea.org/MTCD/publications/PDF/te_1178_prn.pdf, retrieved 16$^{th}$ May 2016

[2] A.R. Junghans, et al., Phys. Lett. B, **670**: 200 (2008)

[3] S. J. Zweben, H. Knoepfel, Phys. Rev. Lett., **35**: 1340 (1975)

[4] R. A. Pitts, et al., Journal of Nuclear Materials, **463**: 39-48 (2013)

[5] A. Shevelev, et al., doi: 10.1063/1.4894038, retrieved 17$^{th}$ May  2016

[6] B. L. Berman, et al., Phys. Rev., **162**: 1098 (1967)

[7] C. H. M. Broeders, et al., Nucl. Eng. Des., **202**: 157 (2000)

[8] F. R. Allum, et al., Nucl. Phys. A, **53** 645 (1964)

[9] H. Naik, et al., Nucl. Phys. A, **916**: 168-182 (2013)

[10] I. Raškinyte˙, et al., in Proc. Int. Conf. on Nuclear Reaction Mechanisms, (Varenna, Italy: Dapnia/SPhN, 2006), DAPNIA-06-147

[11] G. Kim, et al.,  Nucl. Instrum. Methods Phys. Res., Sect. A, **485**: 458-467 (2002)

[12] V. C. Petwal, et al., PRAMANA — journal of physics, **68**: 235 (2007)

[13] M. Gallardo, et al., Phys. Lett. B, **191**: 222-226 (1987)

[14] M. Mattiuzzi, et al., Phys. Lett. B, **364**: 13-18 (1995)

[15] P. Heckman, et al., Phys. Lett. B, **555**: 43 (2003)

[16] Balaram Dey, et al., Phys. Lett. B, **731**: 92-96 (2014)

[17] H. Steinwedel, et al., Z. Naturforsch., **5a**: 413 (1950)

[18] M. Danos, Nucl. Phys., **5**: 23 (1958)

[19] B. L. Berman, At. Data Nucl. Data Tables, **15**: 319-390 (1975)

[20] G. Reffo, Phys. Rev. C, **44**, 814 (1991)

[21] S. Levinger, Phys. Rev., **84**: 43 (1951)

[22] J. S. Levinger, in Nuclear Photo Disintegration (Oxford University Press, Oxford, 1960) p.54

[23] A. Koning et al., TALYS – 1.6 A nuclear reaction program, (2013) p.62

[24] M. Herman et al., EMPIRE – 3.2 Malta modular system for nuclear reaction calculations and nuclear data evaluation, (2013) p.18-20

[25] T. Belgya, et al., Handbook For Calculations of Nuclear Reaction Data, (IAEA, Vienna, RIPL-2,IAEA-TECDOC-1506, 2006), http://www-nds.iaea.org/RIPL-2

[26] V.N. Levkovski, J. Phys., **18**: 361 (1974)

[27] J. S. Wang et al., Eur. Phys. J. A, **7**: 355-360 (2000)

[28] https://www-nds.iaea.org/exfor/exfor.htm, retrieved 4th December 2015

[29] Akito Takahashi, et al., Fusion Engineering and Design, **9**: 323 (1989)

[30] http://www.nndc.bnl.gov/chart/chartNuc.jsp, retrieved 2nd November 2015

[31] https://www-nds.iaea.org/photonuclear/, retrieved 4th December 2015

[32] http://www.nndc.bnl.gov/sigma/index.jsp?dontshow=nn.6&as=9&lib=jendl3.3&nsub=20040, retrieved 4th December 2015

[33] Grady Hughes, Progress in Nuclear Science and Technology, **4**: 454-458 (2014)

[34] X-5 Monte Carlo Team, MCNP-A General Monte Carlo N-Particle Transport Code, Version 5, (2000) 1

[35] A. Fassò, et al., Advanced Monte Carlo for Radiation Physics, Particle Transport Simulation and Applications, in Proceedings of the Monte Carlo 2000 Conference, edited by A. Kling, et. al., (Lisbon, 2000) 159-164

[36] P. K. Sahani et al., Indian J. Pure Appl. Phys, **50**: 863-866 (2012)

[37] Boubaker Askri, Nucl. Instrum. Methods Phys. Res., Sect. B, **360**: 1-8 (2015)